\documentclass[nobibnotes,nofootinbib,superscriptaddress]{revtex4}
\usepackage{amsmath,amssymb,bm}
\usepackage{graphicx}
\usepackage{pstricks,pst-node,pst-text,pst-3d}
\usepackage{datetime}

\newcommand{\beeq}{\begin{eqnarray}}
\newcommand{\eeeq}{\end{eqnarray}}
\newcommand{\be}{\begin{equation}}
\newcommand{\ee}{\end{equation}}
\newcommand{\bea}{\begin{array}}
\newcommand{\eea}{\end{array}}

\newcommand{\eq}{&=&}

\def\qbar{\overline{q}}
\def\pbar{\overline{p}}

\def\ubar{\overline{u}}
\def\dbar{\overline{d}}

\def\half{\textstyle{\frac{1}{2}}}

\begin{document}
\title{\bf Drell-Yan process at forward rapidity at the LHC}

\author{Krzysztof Golec-Biernat}\email{golec@ifj.edu.pl}
\affiliation{Institute of Physics, University of Rzesz\'ow, Rzesz\'ow, Poland}
\affiliation{Institute of Nuclear Physics Polish Academy of Sciences, Cracow, Poland}
\author{Emilia Lewandowska}\email{emilia.lewandowska@ifj.edu.pl}
\affiliation{Institute of Nuclear Physics Polish Academy of Sciences, Cracow, Poland}
\author{Anna M. Sta\'sto}\email{astasto@phys.psu.edu}
\affiliation{Penn State University, Physics Department, University Park, PA 16802, USA}
\affiliation{RIKEN Center, Brookhaven National Laboratory, Upton, NY 11973, USA}
\affiliation{Institute of Nuclear Physics Polish Academy of Sciences, Cracow, Poland}

\begin{abstract}
We analyze the Drell-Yan lepton pair production at forward rapidity at the Large Hadron Collider.
Using the dipole framework for the computation of the cross section we find a 
  significant suppression  in comparison to the collinear factorization formula due to saturation effects in the dipole cross section.
 We develop a twist expansion in powers of $Q_s^2/M^2$ where $Q_s$ is the saturation scale and $M$ the invariant mass of the produced lepton pair. For the nominal LHC energy the leading twist description is sufficient down to masses of $6 \; {\rm GeV}$. Below that value the higher twist terms give a significant contribution. 
 \end{abstract}

\maketitle

\section{Introduction}

The Large Hadron Collider (LHC) opens a new kinematic regime at high energies. In  this regime QCD evolution leads to the   fast growth of the gluon density. At these high densities   it is possible that the   novel phenomena related to the nonlinear dynamics of the gluon fields will occur. Drell-Yan production is a unique process which offers high sensitivity to the parton distribution in the hadron. It is one of the few processes in hadron-hadron collisions where the  collinear factorization  has been rigorously  proven \cite{Collins:1981tt,Collins:1983ju,Collins:1985ue,Collins:1988ig}.   Within this framework, the NLO calculations have been performed in \cite{Altarelli:1978id,Altarelli:1979ub,KubarAndre:1978uy}, and later on up to NNLO accuracy in \cite{Matsuura:1988sm,Hamberg:1990np,Blumlein:2005im}.  In the collinear factorization approach the DY process is viewed as the fusion of the quark and antiquark  which produces a virtual (timelike) photon. 

One can view the same process alternatively in the rest frame of one of the hadrons. The quark (typically valence one) from the fast incoming projectile  interacts with the color field of the target hadron, and emits the virtual photon.  The photon then decays producing lepton pair which moves into the region of forward rapidity (with respect to the incoming projectile).   The original description of the DY in the dipole picture has  been proposed  
in  \cite{Brodsky:1996nj,Kopeliovich:2000fb} with details of the calculations presented in
\cite{Kopeliovich:1998nw}. This process has been also reexamined in \cite{Betemps:2001he} and later on in \cite{Gelis:2002fw,Gelis:2006hy}
within the framework of the Color Glass Condensate.  This formulation is applicable to the very forward production,  when the fractions of the longitudinal momenta of the incoming partons are very different (see discussion in the next section).  This approach has been very useful as one can easily incorporate the higher twist effects due to the multiple scattering of dipole off the target field.  When  the energy is high, one of the momentum fractions of the incoming partons is very small, and as a result one is probing potentially dense gluon fields in the target.
This means that the multiple scatterings have to be taken into account. Consequently the parton evolution should be modified by inclusion of higher twist terms, i.e. terms which are subleading  in the expansion of $Q_s(x)^2/M^2$ where $Q_s(x)$ is the saturation scale characterizing the dense gluon field in the target, and $M^2$ is the invariant mass squared of the produced Drell-Yan pair.

The twist expansion method for analyzing the deep inelastic process has been constructed in \cite{Bartels:2000hv} for the case of the
dipole model based on the idea presented in \cite{Golec-Biernat:1998js}. There, a  systematic expansion in powers of $Q_s(x)^2/Q^2$
was performed  and terms up to twist $4$ were extracted analytically  both for the longitudinal and transverse structure function.
The analysis was further extended to the dipole model which included the DGLAP evolution corrections \cite{Bartels:2009tu}.

In this paper we develop a twist expansion for the Drell-Yan process in a similar approach to that presented in \cite{Bartels:2000hv} for DIS.
Unlike the DIS case however, due to the presence of the second hadron (projectile) the expansion involves additional dependence on the structure function of that fast moving hadron. As a result the obtained expressions still contain the integrals over the longitudinal momentum fractions which involve the structure function.  Using this formalism we evaluate the twist $2$ and twist $4$ contributions to the DY process
as a function of the DY mass. We find that for the nominal LHC energy $14 \; {\rm TeV}$, the leading twist $2$ contribution coincides with the all-twist result for masses down to about $M=6 \;{\rm GeV}$. Below that value resummation of the higher twists contributions is necessary.
We have performed the analysis for both the transverse and longitudinal components and find that the twist 4 contributions have different signs in both cases (for ver low masses), however we do not observe the cancellations of the type found in the case of the DIS process \cite{Bartels:2000hv}.

\section{Drell-Yan cross section}

\begin{figure}[t]
\begin{center}
\includegraphics[width=12cm]{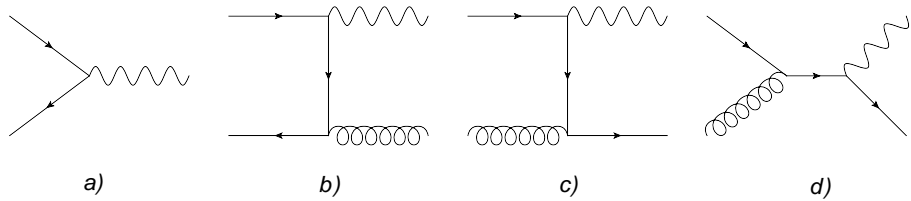}
\caption{The leading and next-to-leading order diagrams for the Drell-Yan production. The diagrams
c) and d) are enhanced in the small-$x$ limit due to a strongly rising gluon distribution.}
\label{fig:dy1}
\end{center}
\end{figure}

In the lowest approximation, the Drell-Yan lepton pair of mass $M$ is produced form annihilation of two quarks of the same flavour $f$
from the colliding hadrons: ~$q_f\qbar_f\to\gamma^*\to l^+l^-$, see Figure \ref{fig:dy1}a. 
In the collinear factorization approach,  the leading order (LO) Drell-Yan cross section is given by
\be\label{eq:1}
\frac{d^2\sigma^{LO}}{dM^2\,dx_F}=\frac{4\pi\alpha^2_{em}}{3N_cM^4}
\frac{x_1x_2}{x_1+x_2}\sum_f e_f^2
\left\{
q_f(x_1,M^2)\,\qbar_f(x_2,M^2)\,+\,\qbar_f(x_1,M^2)\,q_f(x_2,M^2) 
\right\} \; ,
\ee
where  $\alpha_{em}$ is the fine structure coupling constant, $N_c=3$ is the number of quark colors, $q_f/\qbar_f$ are quark/antiquark distributions in the colliding hadrons computed  at the factorization scale $\mu^2=M^2$ and $x_{1,2}$ are the light-cone momentum fractions  of the quarks entering the scattering. 
In the LO approximation, the energy-momentum conservation at the photon vertex,
$(x_1 p + x_2 \pbar)^2=M^2$, leads to the following relation
\be\label{eq:2}
x_1x_2={M^2}/{s}\equiv\tau\ \; , 
\ee
where $s=(p+\pbar)^2=2p\cdot\pbar$ is the   center-of-mass energy squared of the colliding hadrons.
Introducing the Feynman variable of the lepton pair,
$x_F=x_1-x_2$, one can easily find
\be\label{eq:3}
x_1=\half(\sqrt{x_F^2+4\tau}+x_F)\,,~~~~~~~~~~~
x_2=\half(\sqrt{x_F^2+4\tau}-x_F)\,.
\ee

In the next-to-leading order (NLO) approximation, additional emission of a parton (quark or gluon) into the final state  has to be 
taken into account. This is shown by the diagrams in Figure \ref{fig:dy1}b\,-\ref{fig:dy1}d. Because of the emission, the quark
entering the photon vertex carries a fraction $z<1$ of the original parton momentum. Thus, the energy-momentum conservation
at the photon vertex, e.g. $(x_1p+z\,x_2\pbar)=M^2$, gives now
\be\label{eq:3a}
x_1x_2={\tau}/{z}\,,
\ee
and the parton momentum fractions take the form
\be\label{eq:3b}
x_1=\half(\sqrt{x_F^2+4(\tau/z)}+x_F)\,,~~~~~~~~~~~
x_2=\half(\sqrt{x_F^2+4(\tau/z)}-x_F)\,.
\ee
From the parton model conditions, $x_{1,2}<1$, we find that $z>z_{min}=\tau/(1-x_F)$. The NLO correction to the Drell-Yan cross section,
proportional to the strong coupling constant $\alpha_s$, is given in the $\overline{\rm MS}$ factorization scheme by 
\cite{Altarelli:1978id,Altarelli:1979ub,KubarAndre:1978uy}
\beeq\nonumber
\label{eq:3c}
\frac{d^2\sigma^{NLO}}{dM^2\,dx_F}&=&\frac{4\pi\alpha^2_{em}}{3N_c M^4}\,\frac{\alpha_s(M^2)}{2\pi}
\int_{z_{min}}^1\!\!\!dz\,\frac{x_1x_2}{x_1+x_2}
\sum_f e_f^2\, \big\{q_f(x_1,M^2)\,\qbar_f(x_2,M^2)\,D_q(z)
\\\nonumber
\\
&&~~~~~~~~~~~~~~~~~~~~~~~~~~~~~+\,g(x_1,M^2)\left[q_f(x_2,M^2)+\qbar_f(x_2,M^2)\right] D_g(z)\,+\,(x_1\leftrightarrow x_2)
\big\} \; ,
\eeeq
where $x_{1,2}$ are given by relations (\ref{eq:3b}) and $g$ is a gluon distribution. 
The coefficient functions $D_{q,g}$ are of the form
\beeq\label{eq:3d}
D_q(z)\eq C_F\left\{
4(1+z^2)\left(\frac{\ln(1-z)}{1-z}\right)_{+} \!-\, 2\frac{1+z^2}{1-z}\ln z +\delta(1-z)\left(\frac{2\pi^2}{3}-8\right)
\right\} \; ,
\\\nonumber
\\
D_g(z)\eq T_R\left\{
(z^2+(1-z)^2)\ln\frac{(1-z)^2}{z}+\frac{1}{2}+3z-\frac{7}{2}z^2
\right\} \; ,
\eeeq
with $C_F=4/3$, $T_R=1/2$ and the standard $(+)$ prescription to regularize soft gluon   emission singularity at $z=1$. Thus,  the DY cross section in the collinear approximation up to ${\cal{O}}(\alpha_s)$ is the sum
\be\label{eq:3e}
\frac{d^2\sigma^{col}}{dM^2\,dx_F}=\frac{d^2\sigma^{LO}}{dM^2\,dx_F}+\frac{d^2\sigma^{NLO}}{dM^2\,dx_F} \; .
\ee
The higher order (NNLO) corrections \cite{Matsuura:1988sm,Hamberg:1990np,Blumlein:2005im}, proportional to $\alpha_s^2$, lead to much more complicated formulae for the coefficient functions and we will not provide them here.

\section{$pp$ versus $p\pbar$ scattering}

\begin{figure}[t]
\begin{center}
\includegraphics[width=9cm]{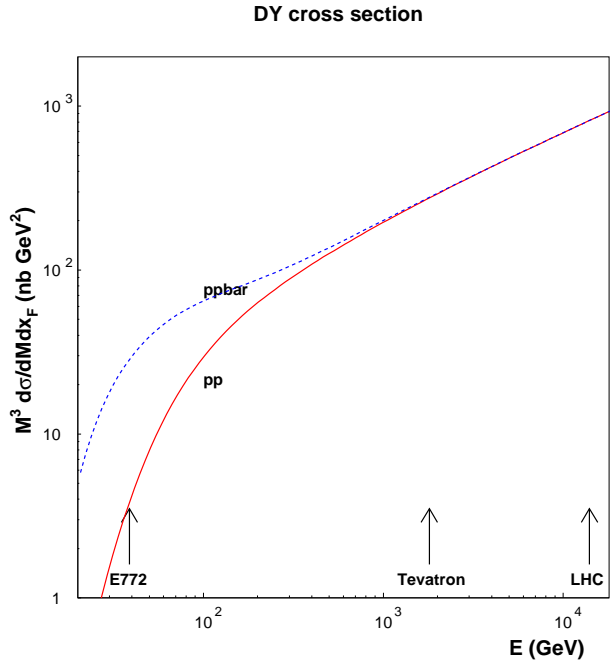}
\caption{The DY cross section in  the collinear approach with CTEQ6.6M parton distributions 
as a function of center-of-mass energy $E=\sqrt{s}$ at fixed $x_F=0.15$ and $M=8{\rm ~GeV}$ 
for $pp$ and $p\pbar$ scattering.}
\label{fig:0}
\end{center}
\end{figure}

We wish to compare the collinear factorization DY cross sections at the two presently operating colliders, the
Femilab Tevatron which scatters $p\pbar$ and the CERN LHC which collides $pp$. Going from the proton into  to the antiproton beam
we have to interchange the quark distributions: $u \leftrightarrow \ubar$, $d \leftrightarrow \dbar$ etc.. 

Such a replacement will not affect
the singlet quark distribution in the second term in eq.~(\ref{eq:3d}) but it will modify the terms with the product of the quark
distributions. For example, for two flavours, in the LO approximation
\be\label{eq:4}
d\sigma^{LO}_{pp}\sim e^2_u\left\{u(x_1)\ubar(x_2)+\ubar(x_1)u(x_2)\right\}+
                 e^2_d\left\{d(x_1)\dbar(x_2)+\dbar(x_1)d(x_2)\right\},
\ee
but in the $p\pbar$ case we find
\be\label{eq:5}
d\sigma^{LO}_{p\pbar}\sim e^2_u\left\{u(x_1)u(x_2)+\ubar(x_1)\ubar(x_2)\right\}+
                 e^2_d\left\{d(x_1)d(x_2)+\dbar(x_1)\dbar(x_2)\right\}.
\ee
Both cross sections coincide if the momentum fraction  $x_2$  is small. In such a case, to a good approximation,  a sea quark is involved in the scattering since
\be
u(x_2)\simeq \ubar(x_2)\,,~~~~~~~~~~~~~~~d(x_2)\simeq \dbar(x_2)\,.
\ee 
This is illustrated in Fig.~\ref{fig:0} where we show the DY cross section  for $pp$ and $p\pbar$ scattering as a function of the center-of-mass energy of colliding particles $E$ at fixed $x_F=0.15$ and $M=8{\rm ~GeV}$. 
For $M\ll E$, $x_1\approx x_F\sim 1$ and 
$x_2\approx \tau/x_F\ll 1$, thus the two cross sections coincide. We use the NLO CTEQ6.6M parton distribution functions  \cite{Nadolsky:2008zw} for this comparison.

\section{Small $x$ limit}

\begin{figure}[t]
\begin{center}
\includegraphics[width=12cm]{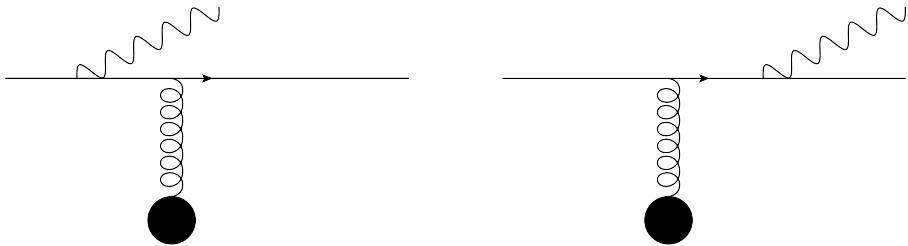}
\caption{The  enhanced diagrams from Fig.~\ref{fig:dy1} in the proton rest frame.}
\label{fig:dy2}
\end{center}
\end{figure}

The small $x$ or high energy limit of the DY process means that
dilepton mass is much smaller than the center-of-mass energy of colliding particles, $M\ll \sqrt{s}$.
When $x_1\sim 1$ we  have 
\be\label{eq:9}
x_2=\frac{M^2}{s\, x_1}\ll 1\,,
\ee
i.e. in the parton model, a fast quark (or antiquark) with the momentum fraction $x_1$ annihilates with a slow antiquark (or quark) with the momentum fraction $x_2$. In such a case, the diagrams in  Figure~\ref{fig:dy1}c\,-\ref{fig:dy1}d, 
with a slow gluon, 
are particularly enhanced due to the strongly rising 
gluon distribution in the small-$x$ limit. Now, a difficult task of small-$x$ resummation
arises \cite{Marzani:2008uh}, which touches the problem of unitarity corrections to the standard, linear QCD evolution equations.

One can reformulate this  problem
in the rest frame of one of the  protons, which acts as a target. In this frame, 
the diagrams in Figure~\ref{fig:dy1}c\,-\ref{fig:dy1}d
 can be interpreted as the lowest order description of the process in which the fast quark scatters
off a soft color field of the target  with emission of a massive photon before or after
the scattering,   see Fig.~\ref{fig:dy2}. The photon subsequently decays into a pair of leptons.

The cross section for radiation of a virtual photon from the fast quark of flavour $f$,  which takes a fraction $z$ of the radiating  quark energy,  is given by \cite{Kopeliovich:2000fb}
\be\label{eq:10}
\sigma_{T,L}^f(qp\to\gamma^*X)=\int d^2r\,
W_{T,L}^f(z,r,M^2,m_f)\,\sigma_{qq}(x_2,zr) \; ,
\ee
where $T,L$ denotes virtual photon polarisation, transverse or longitudinal, respectively.
Here $r$ is the photon-quark transverse separation and $W_{T,L}^f$ are equal to \cite{Brodsky:1996nj,Betemps:2001he}
\beeq\label{eq:11}
W_{T}^f\eq\frac{\alpha_{em}}{\pi^2}\left\{
[1+(1-z)^2]\,\eta^2K_1^2(\eta r)+m_f^2\,z^4K_0^2(\eta r)
\right\} \; ,
\\\label{eq:12}
W_{L}^f\eq\frac{2\alpha_{em}}{\pi^2}\,M^2(1-z)^2K_0^2(\eta r) \; ,
\eeeq
where $K_{0,1}$ are Bessel-McDonald functions, $m_f$ is quark mass and 
$\eta^2=(1-z)M^2+z^2m_f^2$.

The quantity $\sigma_{qq}$ in eq.~(\ref{eq:10}) is a  dipole cross section
known from DIS scattering at small Bjorken-$x$ \cite{Nikolaev:1990ja}. 
It was determined from fits
to HERA data on the proton structure function $F_2$ at small $x$ under different assumptions, e.g. assuming 
phenomenological form with a saturation scale $Q_s^2(x)=(x/x_0)^{-\lambda}$
\cite{Golec-Biernat:1998js,Golec-Biernat:1999qd}:
\be\label{eq:13}
\sigma_{qq}(x,r)=\sigma_0\left\{1-\exp(-r^2Q_s^2(x)/4)\right\}\,.
\ee
Substituting this   dipole cross section into eq.~(\ref{eq:10}) one can test predictions on the DY cross section in which parton saturation effects are taken into account.
The final form of the DY cross section for the forward dilepton production is
found after taking into account the incoming quark distribution in the proton
\be\label{eq:16a}
\frac{d^2\sigma^{DY}_{T,L}}{dM^2\,dx_F}=\frac{\alpha_{em}}{6\pi M^2}\frac{x_1}{x_1+x_2}\,\sum_fe_f^2
\int_{x_1}^1\frac{dz}{z^2}
\left[q_f\!\left(\frac{x_1}{z},M^2\right)+\qbar_f\!\left(\frac{x_1}{z},M^2\right)\right]
\sigma_{T,L}^f(qp\to\gamma^*X)\,.
\ee
The expression in the squared brackets under the integral is proportional to the LO contribution 
of flavour $f$ to the $F_2$ proton structure function, $F_2^f=e_f^2\,x(q_f+\qbar_f)$.
Thus, the final formula reads
\be\label{eq:16}
\frac{d^2\sigma^{DY}_{T,L}}{dM^2\,dx_F}=\frac{\alpha_{em}}{6\pi M^2}\frac{1}{x_1+x_2}\,\sum_f
\int_{x_1}^1\frac{dz}{z}\,F_2^f\!\left(\frac{x_1}{z},M^2\right)\,
\sigma_{T,L}^f(qp\to\gamma^*X)\,.
\ee

A similar expression was found in \cite{Brodsky:1996nj} by changing the variables $r$ and $z$ to 
$\rho=z r$ and $\alpha={(1-z)}/{z}$.
The new variable $\rho$  has the interpretation of 
a size of a quark pair while $\alpha$ is  a quark/antiquark longitudinal momentum fraction
with respect to the photon momentum.

\begin{figure}[t]
\begin{center}
\includegraphics[width=10cm]{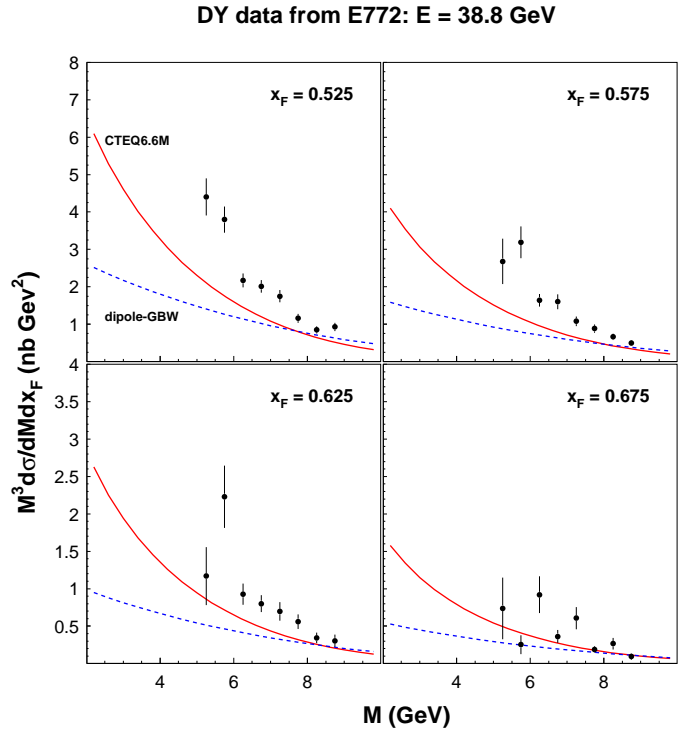}
\caption{The DY cross section from the collinear (solid lines) and dipole (dashed lines) approaches
as a function of the dilepton mass $M$ against the  E772 collaboration data. The CTEQ6.6M parton distributions are used in the collinear case.}
\label{fig:1}
\end{center}
\end{figure}

\section{Predictions for the LHC}

In  Figure \ref{fig:1} we present a comparison of the results from the collinear factorization
formula (\ref{eq:1}) and the dipole formula (\ref{eq:16}) against the data from the Fermilab E772 collaboration \cite{McGaughey:1994dx}. We use the next-to-leading order (NLO) CTEQ6.6M parton distributions \cite{Nadolsky:2008zw} 
for the collinear formula (solid lines)  while the GBW parameterisation \cite{GolecBiernat:2006ba} 
of the dipole cross section was used in the dipole formula.
In eq.~(\ref{eq:16}), there are three massless light quarks and charm quark with mass $m_c=1.4~{\rm GeV}$. 
For the energy $\sqrt{s}={38.8}~{\rm GeV}$ and the indicated values
of $M$ and $x_F$, the fraction of the slow parton momentum $x_2\approx 0.01 - 0.1$, which is slightly beyond the  applicability of the dipole formula. We observe that the data are above the results from both the dipole  and  collinear factorization approaches. A similar result was found for 
the NLO MSTW08 parton distributions \cite{Martin:2009iq}.

\begin{figure}[t]
\begin{center}
\includegraphics[width=8cm]{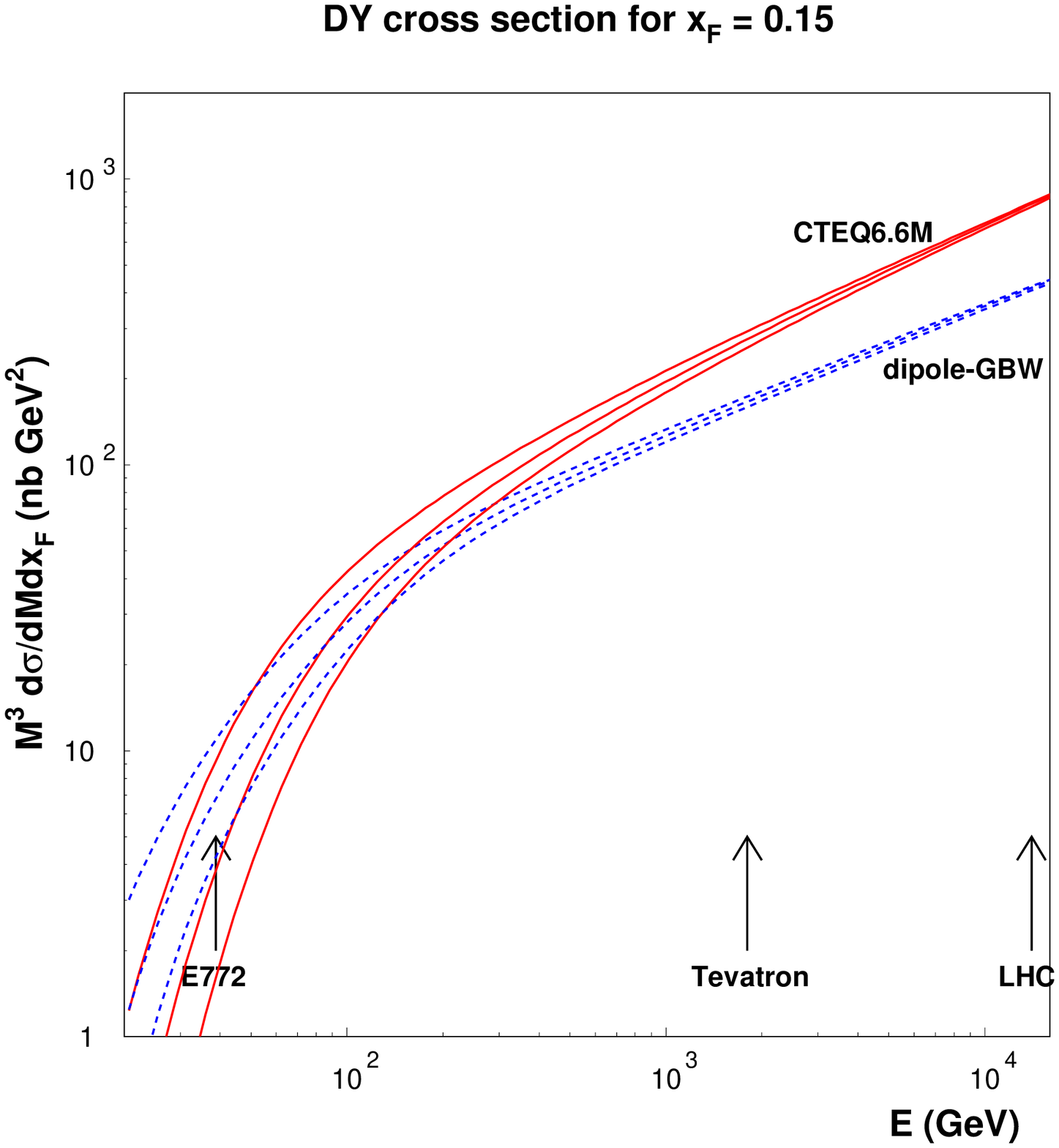}
\includegraphics[width=8cm]{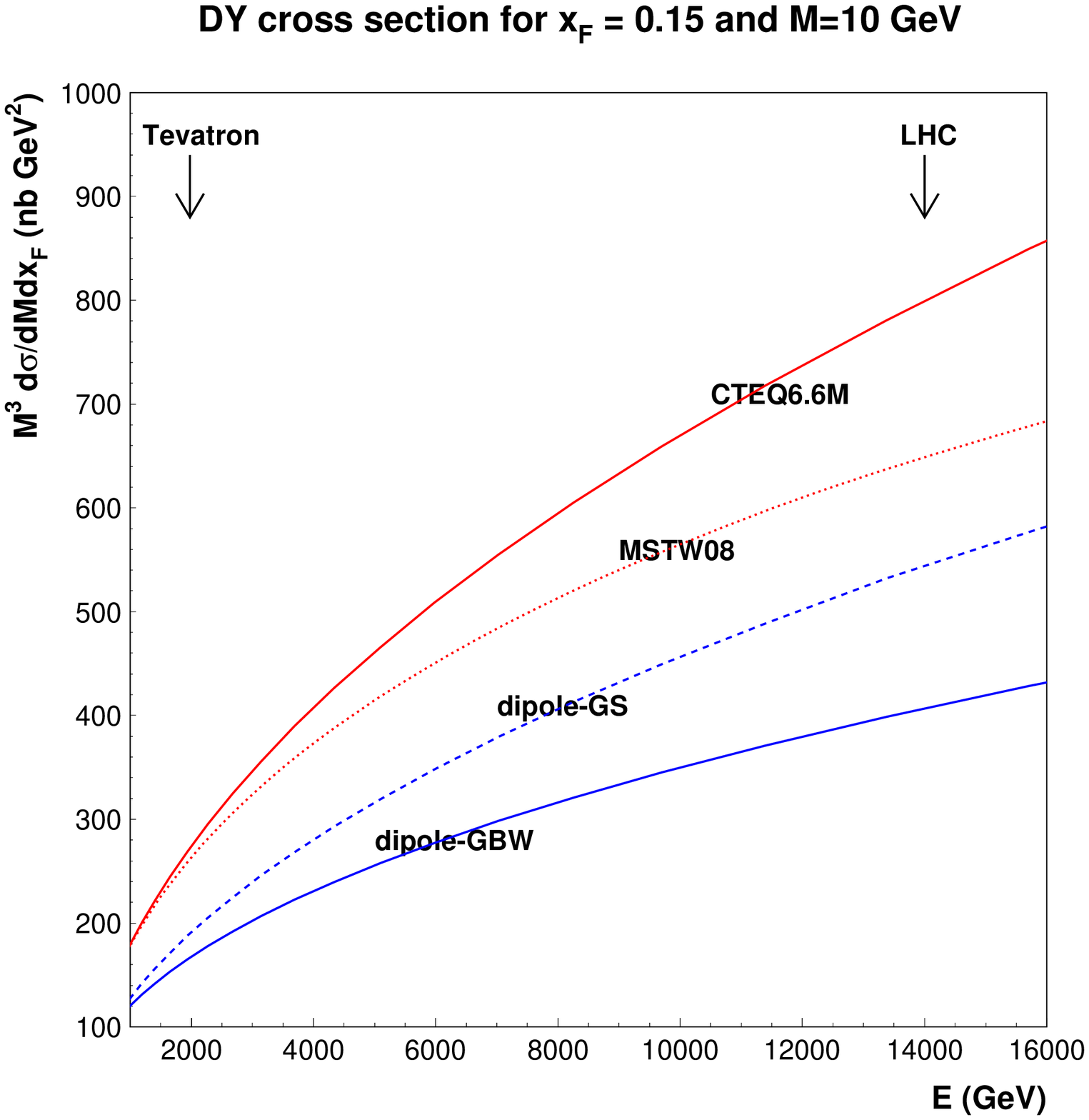}
\caption{Left: the DY cross section from the collinear (solid lines) and dipole (dashed lines) approaches as a function of energy $E=\sqrt{s}$ for fixed
$x_F=0.15$ and dilepton mass $M=6,8,10~{\rm GeV}$ (from top to bottom). Right: the same
for $M=10~{\rm GeV}$ and the indicated parameterizations of the parton distributions and dipole cross section.
}
\label{fig:2}
\end{center}
\end{figure}

In Figure \ref{fig:2} (left) we present predictions for the DY cross section 
as a function of the center-of-mass energy $E=\sqrt{s}$ at fixed $x_F=0.15$ and dilepton
mass $M=6,8,10~{\rm GeV}$. 
For the collinear factorisation results we use the  CTEQ6.6M parton distributions.
In the figure on the right, we show the same results for $M=10~{\rm GeV}$
in a more detailed way, using  the linear scales. We additionally show the collinear factorization results
for the MSTW08 parton distributions and the dipole approach results with two parameterizations of the dipole cross section: 
GBW \cite{Golec-Biernat:1998js} and GS \cite{GolecBiernat:2006ba}. In the latter parameterization the 
DGLAP evolution of the dipole cross
section for small dipole sizes is built in. We also analysed  the Color Glass Condensate parametrization \cite{Soyez:2007kg}, finding results very close to the GBW curves.
At the LHC energy, the fraction $x_2\approx 3\cdot 10^{-6}$ and we are really in the small-$x$ domain which has not been explored experimentally yet at the hard scale given by the invariant mass of the DY lepton pair. Thus, the presented results are only extrapolations. Nevertheless,   
we clearly see that  saturation effects encoded in the dipole approach give results which are systematically below 
the collinear factorization predictions. At the LHC energy, in the most extreme case, the suppression of the
DY cross section due to saturation effects can be as large as a factor of three.

We are looking forward to the experimental verification of this result.

\section{Twist expansion of the dipole formula}

We will analyze the dipole DY cross section (\ref{eq:16}) from the point of view of the twist expansion in positive powers of the ratio $Q_s^2(x_2)/M^2$, where $Q_s$ is the $x$-dependent saturation scale.
 We  utilize the Mellin transform, using the methods elaborated  in \cite{Bartels:2000hv,Bartels:2009tu}.  The twist analysis for  the Drell-Yan production is slightly more complicated than in the case of the structure functions. This is due to the fact that the integral in $z$ cannot be performed analytically. This is because it involves the structure function $F_2(x_1/z)$ which is given as  a parametrization to the experimental data.  In order to compute the twist expansion we will present two methods. In the first one we will expand the integrand around the $z=1$ point and perform the integrals analytically. This will give us approximate, analytical expressions
 for higher twists. In the second method we leave the integrals in $z$ and devise a method to extract the twist contributions numerically. This method gives exact results for the twist expansion.

The standard definition of the Mellin transform of a function $f(r^2)$ reads
\be
\phi(\gamma)=\int_0^{\infty} \frac{dr^2}{r^2}\,(r^2)^{-\gamma} f(r^2) \;,
\label{eq:Mellin1}
\ee
while the inverse transform is given by
\be
f(r^2) = \int_{c-i\infty}^{c+i\infty} \frac{d\gamma}{2\pi i}\,(r^2)^{\gamma}\phi(\gamma) \; ,
\label{eq:Mellin2}
\ee
where $c$ is  a real number which has to be taken in interval $(a,b)$  such that 
the integral (\ref{eq:Mellin1}) is absolutely convergent for $a<{\mathcal Re}\,\gamma<b$.

We start the twist  analysis from the transverse part of the cross section, and we assume that  quarks are massless. The part of the DY cross section which corresponds to the transverse polarization of the photon   (\ref{eq:16}) reads explicitly 
\be
\frac{d^2\sigma^{DY}_T}{dM^2\,dx_F}=\frac{\alpha_{em}^2}{6\pi^2 M^2}\frac{1}{x_1+x_2}
\int_{x_1}^1\frac{dz}{z}\,\,F_2\!\left(\frac{x_1}{z},M^2\right) \int_0^{\infty} dr^2\, [1+(1-z)^2] \, 
M^2(1-z) \,K_1^2(Mr\sqrt{1-z}) \, \sigma_{q\bar{q}}(x_2,zr) \, .
\label{eq:sigt}
\ee
Using the above definitions of the Mellin transform,  the  dipole cross section can be written as
\be
\sigma_{q\bar{q}} (x,r)\, = \, \int_{c-i\infty}^{c+i\infty}\frac{d\gamma}{2\pi i}\, (r^2)^{\gamma}\int_0^{\infty} \frac{dr'^2}{r'^2}
\,(r'^2)^{-\gamma}\,\sigma_{q\bar{q}} (x,{r'}) \;.
\label{eq:dipolemellin}
\ee
 We will perform the twist analysis using the GBW parametrization of the dipole cross section (\ref{eq:13})
 which is given in a closed analytic form.
 The advantage of this is that we can perform the integral over the dipole sizes analytically.

 The Mellin transform of the GBW cross section reads
 \be
 \sigma_0\, G(\gamma) \equiv \sigma_0
\int_0^{\infty}\,\frac{d\hat{r}^2}{\hat{r}^2}(\hat{r}^2)^{-\gamma}(1-e^{-\hat{r}^2})
=-\sigma_0\,\Gamma(-\gamma) \; ,
 \label{eq:gbwmellin}
 \ee
 and therefore has single poles for all non-negative integer values of $\gamma$.

Using representation (\ref{eq:dipolemellin}) together with the explicit Mellin transform of the GBW cross section  (\ref{eq:gbwmellin}), we can rewrite the above expression as
\beeq\nonumber
 \frac{d^2\sigma^{DY}_T}{dM^2\,dx_F}&=&\frac{\alpha_{em}^2}{6\pi^2 M^2}\frac{1}{x_1+x_2}
\int_{x_1}^1\frac{dz}{z}\,\,F_2\!\left(\frac{x_1}{z},M^2\right) \int_0^{\infty} dr^2\, [1+(1-z)^2]\, M^2(1-z) \, K_1^2(Mr\sqrt{1-z})
\\\nonumber
\\
&\times& \; \sigma_0 \int_{c-i\infty}^{c+i\infty} \frac{d\gamma}{2\pi i} (r^2)^{\gamma} \bigg(\frac{z^2 Q_s^2(x_2)}{4}\bigg)^{\gamma} G(\gamma) \; ,
\eeeq
where the contour of integration over $\gamma$ is chosen so that the constant  $c$ satisfies $0<{{\rm Re} \,c}<1$ and ${\rm Im}\, c =0$.
We can now perform the integration over the dipole size $r$. To this aim we define
\be
\widetilde{H}_T(\gamma)\; \equiv \;
\int_0^{\infty} d\tilde{r}^2 K_1^2({\tilde{r}}) (\tilde{r}^2)^{\gamma} 
\; = \; \frac{\sqrt{\pi}\Gamma(\gamma)\Gamma(1+\gamma)\Gamma(2+\gamma)}{2\Gamma(\frac{3}{2}+\gamma)} \; .
\label{eq:K12mellin}
\ee
The transverse part of the DY cross section  reads therefore
\beeq\nonumber
 \frac{d^2\sigma^{DY}_T}{dM^2\,dx_F}&=&\frac{\alpha_{em}^2\sigma_0}{6\pi^2 M^2}\frac{1}{x_1+x_2}  \int_{c-i\infty}^{c+i\infty} \frac{d\gamma}{2\pi i} \,G(\gamma) \widetilde{H}_T(\gamma)\,\bigg(\frac{ Q_s^2(x_2)}{4M^2}\bigg)^{\gamma} 
 \\\nonumber
\\
&\times&  \int_{x_1}^1\frac{dz}{z}\,\,F_2\!\left(\frac{x_1}{z},M^2\right) [1+(1-z)^2] \left(\frac{z^2}{1-z}\right)^{\gamma} \; .
\label{eq:DYTexpl}
\eeeq
Clearly the poles in the $\gamma$ plane control the behavior in $M^2$.  We need to evaluate the integral over the longitudinal momentum fraction $z$.  Let us introduce the following notation
\be
I_{T,{\gamma}}(x_1,z,M^2) = \frac{1}{z}\,\,F_2\!\left(\frac{x_1}{z},M^2\right) [1+(1-z)^2] (z^2)^{\gamma} \; .
\label{eq:IT}
\ee
We first perform the integral over $z$ analytically by expanding the above expression around  $z=1$. We  assume that $F_2\!\left({x_1}/{z},M^2\right)$ does not have any singularities at $z=1$,
which is corroborated by the expressions for the structure functions at these values of $x_1\sim 1$.

Expanding the integrand and integrating term by term one obtains, 
\be
\int_{x_1}^1 dz \sum_{k\ge 0} I_{T,{\gamma}}^{(k)}(x_1,z=1,M^2)(1-z)^k \left(\frac{1}{1-z}\right)^{\gamma}= \sum_{k\ge 0} I_{T,{\gamma}}^{(k)}(x_1,z=1,M^2) (1-x_1)^{1-\gamma+k}\frac{1}{1-\gamma+k}  \; .
\label{eq:ITexpand}
\ee
Inserting this expansion into eq.~(\ref{eq:DYTexpl}) one obtains a general expression which allows to extract the twists systematically in powers of $(1-x_1)$
\begin{equation}
 \frac{d^2\sigma^{DY}_T}{dM^2\,dx_F}=\frac{\alpha_{em}^2\sigma_0}{6\pi^2 M^2}\frac{1}{x_1+x_2}  \, \sum_{k\ge 0} \, \int_{c-i\infty}^{c+i\infty} \frac{d\gamma}{2\pi i} \,G(\gamma) \widetilde{H}_T(\gamma)\,\frac{I_{T,{\gamma}}^{(k)}(x_1,z=1,M^2)}{1-\gamma+k} \,  (1-x_1)^{1+k} \, \bigg(\frac{ Q_s^2(x_2)}{4M^2(1-x_1)}\bigg)^{\gamma} \, .
 \label{eq:DYTexpansion}
\end{equation}
 
 The twist expansion then corresponds to taking residues of the different poles in $\gamma$
 that appear on the right-hand side of (\ref{eq:DYTexpansion}).
One can also rewrite it to expose a more general structure of this expansion
\be
 \frac{d^2\sigma^{DY}_T}{dM^2\,dx_F} \; = \; \sum_{\gamma_c \ge 1} \;  \sum_{k\ge 1-\gamma_c} (1-x_1)^k \bigg( \frac{Q_s^2(x_2)}{4M^2} \bigg)^{\gamma_c} \, C_{k,\gamma_c}(x_1,\ln Q_s^2/(M^2(1-x_1)),\ln M^2) \; ,
 \label{eq:sumgammack}
\ee
where the first sum is performed over the poles in eq.~(\ref{eq:DYTexpansion}).
The coefficients $C_{k,\gamma_c}$ depend on $\ln Q^2_s/(M^2(1-x_1))$ which reflects the fact that there can be multiple poles, and the $\ln M^2$ dependence comes from the possible dependence of $F_2$ on $M^2$. Since $F_2$ is evaluated at large values of $x_1$ we expect this dependence to be very mild.

\subsection{Twist $2$}

To extract the twist $2$ in powers of $Q_s^2(x_2)/M^2$ we  will  consider the leading term in the expansion of the function $I_T$ around $z=1$. 
Taking this term, the integral over $z$  reads
\beeq\nonumber
\int_{x_1}^1dz\,\, I_{T,{\gamma}}^{(0)}(x_1,z=1,M^2) \left(\frac{1}{1-z}\right)^{\gamma} 
&=&\int_{x_1}^1dz F_2\!\left(x_1,M^2\right) \left(\frac{1}{1-z}\right)^{\gamma} 
\\\nonumber
\\
&=&  F_2\!\left(x_1,M^2\right) (1-x_1)^{1-\gamma} \frac{1}{1-\gamma} \; .
\label{eq:IT0}
\eeeq
Therefore integral over $z$ gives the single pole in the $\gamma$ plane.
The function $G(\gamma)=-\Gamma(-\gamma)$ also has a single pole, which together gives the  double pole in $\gamma=1$. The higher order terms in expansion of (\ref{eq:IT}) will contribute to terms
which are suppressed by powers of $(1-x_1)$ as is evident from eq.~(\ref{eq:sumgammack}). We will discuss these later.

Taking the first term in the expansion in $1-z$, (\ref{eq:IT0}) we obtain  the approximate formula
\begin{equation}
 \frac{d^2\sigma^{DY}_T}{dM^2\,dx_F}\simeq\frac{\alpha_{em}^2\sigma_0}{6\pi^2 M^2}\frac{  F_2\!\left(x_1,M^2\right)}{x_1+x_2} (1-x_1) \int_{c-i\infty}^{c+i\infty} \frac{d\gamma}{2\pi i} \,G(\gamma) \widetilde{H}_T(\gamma)\frac{1}{1-\gamma}\,\bigg(\frac{ Q_s^2(x_2)}{4M^2(1-x_1)}\bigg)^{\gamma}  \; .
\label{eq:dylt1}
\end{equation}
The leading twist extraction amounts to taking the above formula and 
closing the contour to the right and taking the contribution from the pole at $\gamma=1$.
 The  approximate expression for the leading twist is therefore
\be
\Delta_{T,2}^{(0)}\,=\,
\frac{\alpha_{em}^2\sigma_0}{6\pi^2 M^2}\frac{  F_2\!\left(x_1,M^2\right)}{x_1+x_2}\times 2  \frac{ Q_s^2(x_2)}{4M^2} \left[\frac{4}{3}\gamma_E-1+\frac{2}{3}\psi(\frac{5}{2})-\frac{2}{3} \ln \frac{ Q_s^2(x_2)}{4M^2(1-x_1)} \right] \; + {\cal O}(1-x_1) \; .
 \label{eq:dytlt2}
\ee
Note that there is an additional dependence on the mass $M^2$ in the function $F_2(x_1,M^2)$.
As $x_1\simeq 1$ this dependence should be very mild and should not affect too much the twist expansion we are performing .
We stress that this expression is an approximate one in the sense that there are subleading contributions coming from the other terms in expression (\ref{eq:IT}).  They will not modify however the logarithmic term as they are regular as $\gamma=1$. More specifically the first subleading contribution to the leading twist comes from
\beeq\nonumber
\label{eq:pole2}
\int_{x_1}^1dz\,\, I_{T,\gamma}^{(1)}(x_1,z=1,M^2) \left(\frac{1}{1-z}\right)^{\gamma} 
&=&\int_{x_1}^1dz (F_2\!\left(x_1,M^2\right)(1-2\gamma)+x_1 F_2'(x_1,M^2)) \left(\frac{1}{1-z}\right)^{\gamma-1} 
\\\nonumber
\\
&=& [F_2\!\left(x_1,M^2\right)(1-2\gamma)+x_1 F_2'(x_1,M^2)] (1-x_1)^{2-\gamma} \frac{1}{2-\gamma} \; .
\eeeq
This contribution is the leading one for the twist 4, but it also will affect the twist 2
as it multiplies the single pole in $\gamma=1$ present in $G(\gamma)$.
It gives the following correction to the leading expression for  twist 2, eq.~(\ref{eq:dytlt2}):
\be
\Delta_{T,2}^{(1)}\,=\,
 -\frac{\alpha_{em}^2\sigma_0}{6\pi^2 M^2}\frac{ [  F_2\!\left(x_1,M^2\right) -x_1F_2'\!\left(x_1,M^2\right) ] }{x_1+x_2}  (1-x_1)\frac{ Q_s^2(x_2)}{4M^2} \times \frac{4}{3} \; ,
 \label{eq:dyltcorr1}
\ee
which is suppressed by one power of $(1-x_1)$.

The above-presented method allows to systematically compute the contributions to the given twist in powers of $(1-x_1)$. The effectiveness of this calculation depends however on the rate of the convergence of the series in $(1-x_1)$. In our  kinematics $x_1$ is about $0.1 - 0.2$, hence this convergence is rather slow. Below we will present another semi-analytical method which allows to compute the leading twist expression exactly.

As the double pole contribution has been evaluated already in (\ref{eq:dytlt2}), the remaining single pole contribution comes from the pole in $G(\gamma)$ multiplying the non-singular in $\gamma=1$ part  of the integral over $z$. It suffices to take the regularized integral
\be
\int_{x_1}^1 dz \frac{I_{T,\gamma=1}(x_1,z,M^2)-I_{T,\gamma=1}^{(0)}(x_1,z=1,M^2)}{1-z} = \int_{x_1}^1 dz \frac{zF_2(\frac{x_1}{z},M^2)(1+(1-z)^2)-F_2(x_1,M^2)}{1-z} \; ,
\ee
which is evaluated at $\gamma=1$.
The integrand in this expression is regular at $z=1$ and thus the integral can be evaluated purely numerically.  The correction to eq.~(\ref{eq:dytlt2}) is therefore
\be
\Delta_{T,2}^{(k>0)}\,=\,
\frac{\alpha_{em}^2\sigma_0}{6\pi^2 M^2}\frac{1}{x_1+x_2} \times \frac{4}{3} \left( \frac{Q_s^2(x_2)}{4M^2}\right) \int_{x_1}^1 dz \, \frac{zF_2(\frac{x_1}{z},M^2)(1+(1-z)^2)-F_2(x_1,M^2)}{1-z} \; ,
\ee
and the complete, exact leading twist expression equals
\be\label{eq:sumtw2t}
\frac{d^2\sigma^{DY(\tau=2)}_T}{dM^2\,dx_F}\, = \, \Delta_{T,2}^{(0)}+\Delta_{T,2}^{(k>0)} \; .
\ee
A similar twist-2 analysis of the longitudinal part of the DY cross section is given in Appendix A.

\begin{figure}[t]
\begin{center}
\includegraphics[width=8cm]{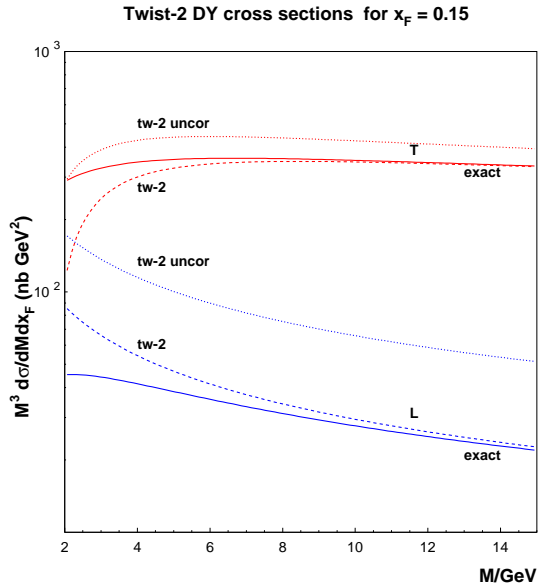}
\caption{The twist-2 contributions, shown by the dashed lines, for the transverse and longitudinal DY cross section at the LHC energy $\sqrt{s}=14~{\rm TeV}$. The solid lines show the all-twist results while the dotted lines
correspond to the approximate twist-2 results given by eqs.~(\ref{eq:dytlt2}) and (\ref{eq:sumtw2la}).
}
\label{fig:1b}
\end{center}
\end{figure}

In Fig.~\ref{fig:1b} we show the comparison of the calculation of the all-twist formulae, 
eq.~(\ref{eq:DYTexpl}) for the transverse and eq.~(\ref{eq:mixedlong}) for the longitudinal parts (shown by the solid lines),  with the approximate leading twist contribution. The approximate twist-2 contributions
$\Delta_{T,2}^{(0)}$ and $\Delta_{L,2}^{(0)}$, eqs.~(\ref{eq:dytlt2}) and (\ref{eq:sumtw2la}) respectively,  are shown by the dotted lines while the exact twist-2
expressions, eq.~(\ref{eq:sumtw2t}) and eq.~(\ref{eq:sumtw2l}), are depicted by the dashed lines.
 We see that the approximate  twist-2 formulae are significantly above the exact result even for lager
valaues of $M$.
This is due to the large value of the subleading terms in the expansion of $(1-x_1)$. Since $x_1 \simeq x_F =0.15$ the higher order terms can be still contributing.
On the other hand, we observe that the exact twist-2 formulae are getting close to the all-twist result in the region $M>6$.

\subsection{Twist $4$}

\begin{figure}[t]
\begin{center}
\includegraphics[width=8cm]{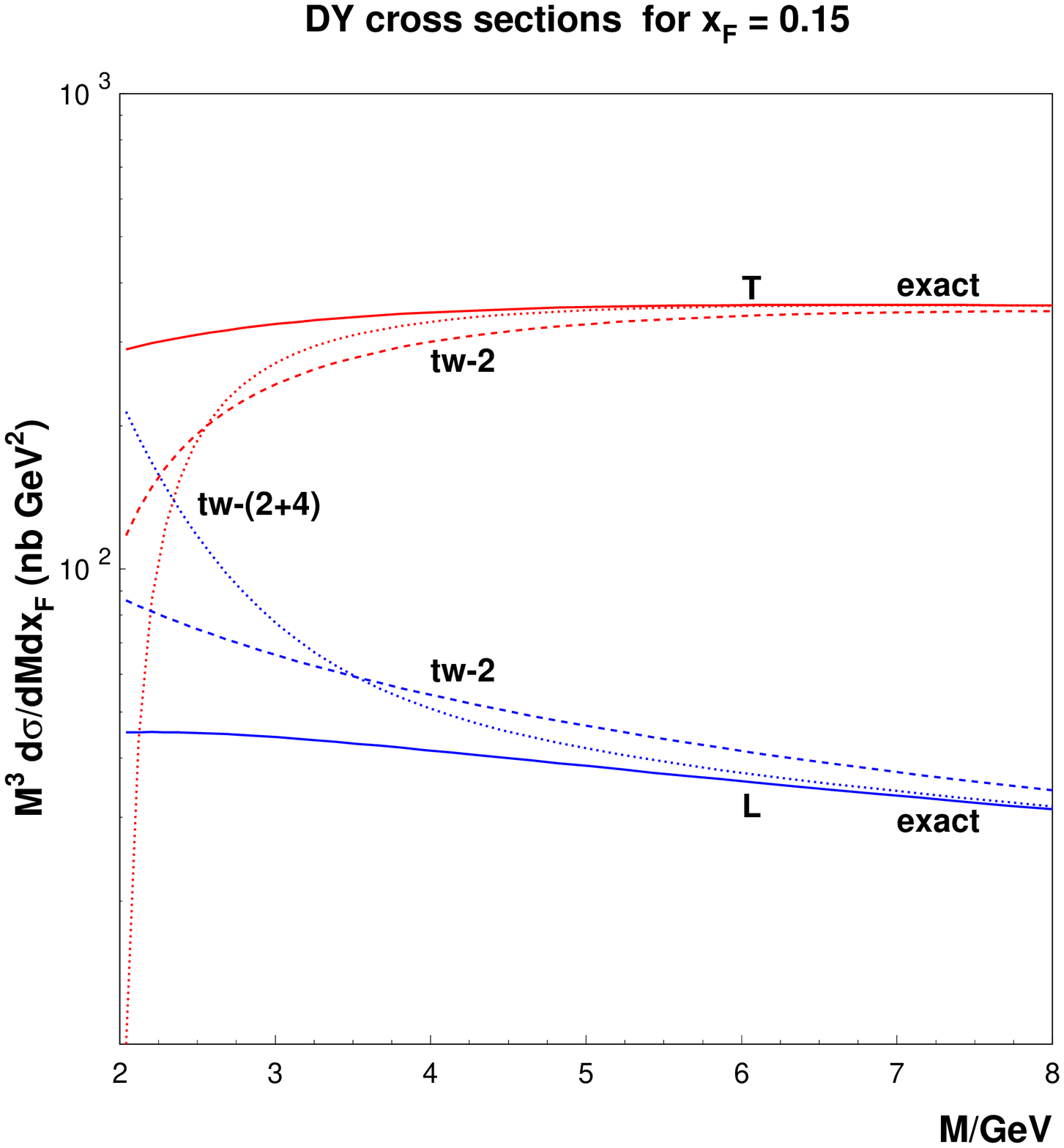}
\includegraphics[width=8cm]{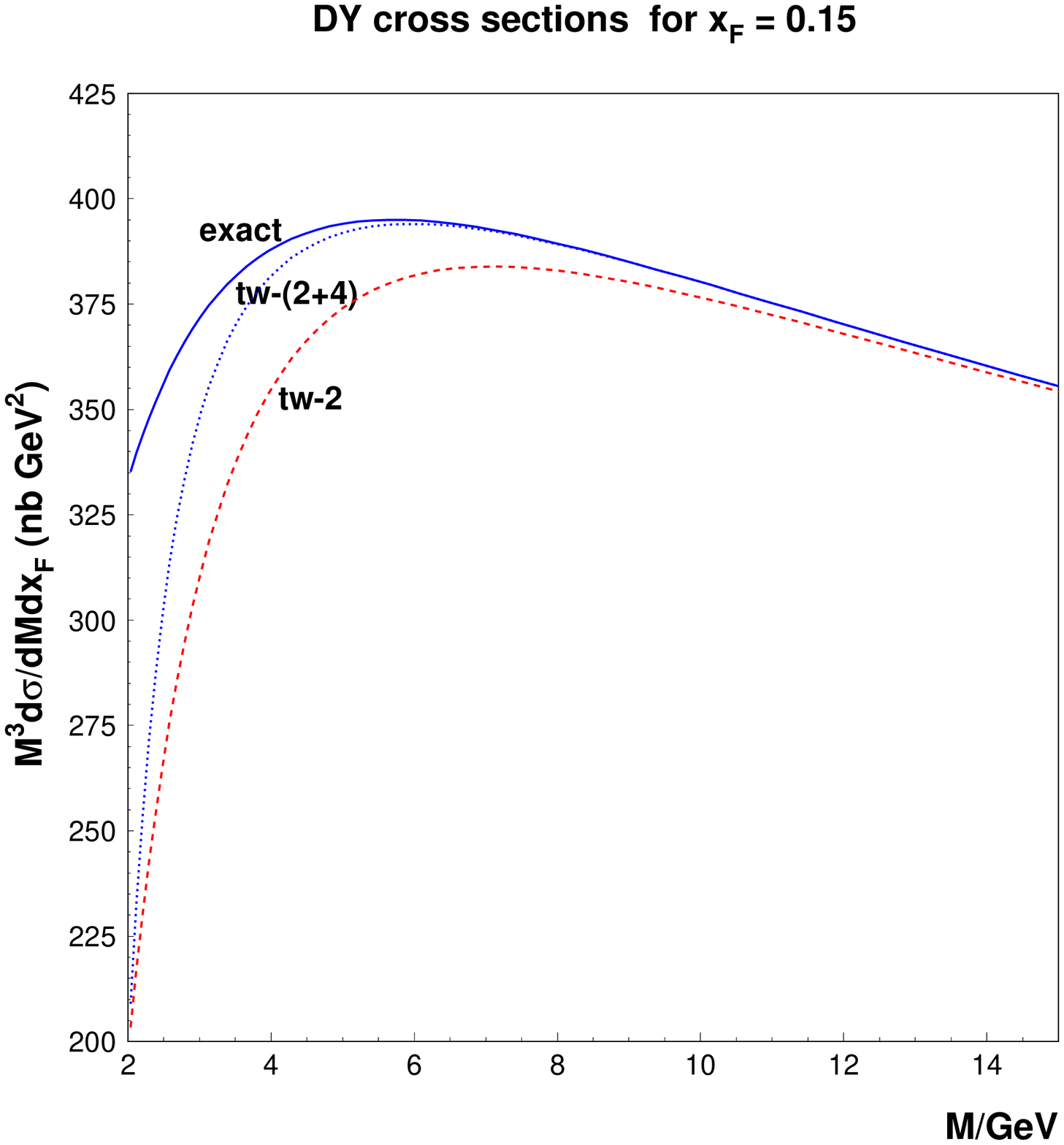}
\caption{Convergence of the twist expansion of the transverse and longitudinal DY  cross sections (left)
and of their sum (right).
The solid lines show the all-twist results while the dashed and dotted lines correspond to the twist-2 and twist-(2+4) contributions, respectively.}
\label{fig:1c}
\end{center}
\end{figure}

In this section we provide  semi-analytical expressions for  the twist $4$ both for the
longitudinal and transverse polarizations of the DY photon. 
We start with the transverse part and use eqs.~(\ref{eq:DYTexpl}) and (\ref{eq:pole2}) together,  which will give part of the twist 4 contribution. Taking the residue at $\gamma=2$ we obtain

\begin{multline}
\Delta_{T,4}^{(1)} = \frac{\alpha_{em}^2\sigma_0}{6\pi^2 M^2}\frac{ 1 }{x_1+x_2}  \bigg( \frac{ Q_s^2(x_2)}{4M^2} \bigg)^2 \times \frac{4}{15} 
\bigg[
F_2(x_1,M^2)\left(  
-63+36\,\gamma_E+18\, \psi(7/2)-18\log \bigg(\frac{ Q_s^2(x_2)}{4M^2(1-x_1)}\bigg)
\right)  
\\
+ x_1F_2^\prime(x_1,M^2)\left(
17-12\, \gamma_E-6\,\psi(7/2)+6\log\bigg(\frac{ Q_s^2(x_2)}{4M^2(1-x_1)}\bigg)
\right) 
\bigg] \; .
\end{multline}
The second part contributing to twist four comes from the contribution of the single pole in $G(\gamma)$ multiplying the finite part in the integral over $z$.  This can be evaluated in the similar manner as before by modifying the integral in order to perform analytical continuation to 
$\gamma=2$.  To this aim we subtract in the integrand the first two terms of the expansion in $(1-z)$
\begin{multline}
\delta {\cal F}(x_1,M^2) = \int_{x_1}^1 dz \frac{ z^3 F_2(\frac{x_1}{z},M^2)(1+(1-z)^2)-F_2(x_1,M^2)-(1-z)[-3 F_2(x_1,M^2) +x_1 F_2'(x_1,M^2)] }{(1-z)^2} \; ,
\end{multline}
which gives convergent integral by definition. Then we need to add the term coming from the lowest order term in expansion in $z=1$ evaluated at $\gamma=2$ which is 
$$
-\frac{F_2(x_1,M^2)}{1-x_1} \; .
$$
 The expression contributing to twist 4 is thus
\begin{equation}
\Delta_{T,4}^{(2)} \; = \; \frac{\alpha_{em}^2\sigma_0}{6\pi^2 M^2}\frac{ 1 }{x_1+x_2}\bigg( \frac{ Q_s^2(x_2)}{4M^2} \bigg)^2 \times (-\frac{8}{5}) \bigg[ \delta {\cal F}(x_1,M^2)-\frac{F_2(x_1,M^2)}{1-x_1} \bigg] \; ,
\end{equation}
where we evaluated the residue coming from the single pole at $\gamma=2$.
The final expression for twist $4$ for the transverse part is therefore
\begin{equation}
\frac{d^2\sigma^{DY(\tau=4)}_T}{dM^2\,dx_F} \; = \; \Delta_{T,4}^{(1)} +\Delta_{T,4}^{(2)}  \; .
\end{equation}

In Fig.~\ref{fig:1c} (left) we show convergence of the twist expansion for the transverse and longitudinal
DY cross sections (left plot) and their sum (right plot). The solid lines show the all-twist results while
the dashed and dotted lines correspond to the twist-2 and twist-(2+4) contributions, respectively.
We see that there is a poor convergence below $M<6~{\rm GeV}$. Both twist contributions are below the exact
results for the transverse part of the cross section while for the longitudinal part they are above the exact
result. This is the reason why the sum of the transverse and longitudinal parts of the twist contributions 
approaches the exact total cross section. This effect is shown in   Fig.~\ref{fig:1c} (right).

\section*{Summary}

In this paper we have investigated Drell-Yan production at forward rapidities and at high energies accessible at the LHC.   We have used the dipole formulation suitable for forward rapidities and used the cross section which incorporates saturation effects.
The comparison with the standard collinear formula shows the suppression of the production cross section for highest energies when using the dipole models with saturation effects. 
 
 Using the dipole formulation with the GBW formula for the dipole model cross section, we have
 constructed the twist expansion for this process.  Unlike the DIS case, where the twist expansion for the GBW model could be performed completely analytically, here we had to resort to the semi-analytical evaluation of the individual terms in the expansion. 
 It was shown that the leading twist is a good approximation to the full result for masses
 of the Drell-Yan pair larger than about $\sim 6 \; {\rm GeV}$. For lower masses the twist expansion quickly becomes divergent and full resummation is necessary.


\begin{acknowledgments}
This work is partially supported by the grants of MNiSW No. 
N202 249235 and the grant HEPTOOLS, MRTN-CT-2006-035505.
A.M.S. is supported by the Sloan Foundation and the DOE OJI grant No. DE - SC0002145.
\end{acknowledgments}

\appendix

\section{Twist decomposition of the  longitudinal part}

The leading twist-2 contribution for the  longitudinal part of the cross section
can be computed in a similar manner as for the transverse part. We use
\be
\widetilde{H}_L(\gamma)\; \equiv \;
\int_0^{\infty} d\tilde{r}^2 K_0^2({\tilde{r}}) (\tilde{r}^2)^{\gamma}\; = \; \frac{\sqrt{\pi}\,\Gamma(1+\gamma)^3}{2\,\Gamma(\frac{3}{2}+\gamma)} \; .
\label{eq:K02mellin}
\ee
After performing the integrals over the transverse coordinate and using the Mellin representation we obtain
\be
 \frac{d^2\sigma^{DY}_L}{dM^2\,dx_F}=\frac{\alpha_{em}^2\sigma_0}{6\pi^2 M^2}\frac{2}{x_1+x_2} \int_{c-i\infty}^{c+i\infty} \frac{d\gamma}{2\pi i} \,G(\gamma) \widetilde{H}_L(\gamma)\,\bigg(\frac{ Q_s^2(x_2)}{4M^2}\bigg)^{\gamma}
\int_{x_1}^1\frac{dz}{z}\,\,F_2\!\left(\frac{x_1}{z},M^2\right) (1-z) \left(\frac{z^2}{1-z}\right)^{\gamma}.
\label{eq:mixedlong}
\ee
Similarly to the transverse case, we can introduce the function
\be
I_L(x_1,z,M^2) = \frac{1}{z}\,\,F_2\!\left(\frac{x_1}{z},M^2\right)  (z^2)^{\gamma} \; .\ee
The leading term coming from the expansion around $z=1$ will thus give
\begin{multline}
\int_{x_1}^1dz\,\, I_L^{(0)}(x_1,z=1,M^2) \left(\frac{1}{1-z}\right)^{\gamma-1} =\int_{x_1}^1dz F_2\!\left(x_1,M^2\right) \left(\frac{1}{1-z}\right)^{\gamma-1} 
=  F_2\!\left(x_1,M^2\right) (1-x_1)^{2-\gamma} \frac{1}{2-\gamma} \; .
\end{multline}
Note that this expression  is not singular at leading pole $\gamma=1$.
Again the leading twist extraction in the longitudinal case therefore amounts to taking the formula
\begin{equation}
 \frac{d^2\sigma^{DY}_L}{dM^2\,dx_F}=\frac{\alpha_{em}^2\sigma_0}{6\pi^2 M^2}\frac{  2 F_2\!\left(x_1,M^2\right)}{x_1+x_2} (1-x_1)^2 \int_{c-i\infty}^{c+i\infty} \frac{d\gamma}{2\pi i} \,G(\gamma) \widetilde{H}_L(\gamma)\frac{1}{2-\gamma}\,\bigg(\frac{ Q_s^2(x_2)}{4M^2(1-x_1)}\bigg)^{\gamma}  \; ,
\label{eq:dyllt1}
\end{equation}
and closing the contour to the right, enclosing the pole at $\gamma=1$. The final result is
\be
\Delta_{L,2}^{(0)}
\simeq\frac{\alpha_{em}^2\sigma_0}{6\pi^2 M^2}\frac{ 2 F_2\!\left(x_1,M^2\right)}{x_1+x_2}  (1-x_1)\frac{ Q_s^2(x_2)}{4M^2} \times \frac{2}{3} \;  + {\cal O}((1-x_1)^2) \; .
 \label{eq:sumtw2la}
\ee
There are no logarithmic corrections because of the single leading pole  coming  only  from  the function $G(\gamma)$. 

Using analogous method as before we can also evaluate the  exact expression for the longitudinal part of the cross section. The integral over $z$ in (\ref{eq:mixedlong})
is well defined for $\gamma=1$ and so to evaluate the exact twist $2$ we can perform this integral directly numerically. 
The exact expression for the twist $2$ contribution to the longitudinal part of the cross section reads
\be
\frac{d^2\sigma^{DY(\tau=2)}_L}{dM^2\,dx_F} = \frac{\alpha_{em}^2\sigma_0}{6\pi^2 M^2}\frac{ 2 }{x_1+x_2}  \frac{ Q_s^2(x_2)}{4M^2} \times \frac{2}{3}  \int_{x_1}^1 dz \, z \, F_2\!\left(\frac{x_1}{z},M^2\right)\; .
\label{eq:sumtw2l}
\ee

Analogously we evaluate the twist-4 contribution to the longitudinal part. The first part of this contribution  is obtained by taking the residue at $\gamma=2$ of expression (\ref{eq:dyllt1})
\begin{equation}
\Delta_{L,4}^{(1)} \; = \; 
\frac{\alpha_{em}^2\sigma_0}{6\pi^2 M^2}\frac{ 2F_2(x_1,M^2) }{x_1+x_2} \bigg( \frac{ Q_s^2(x_2)}{4M^2} \bigg)^2 \times \bigg( -\frac{16}{15} \bigg) \bigg[ -3+2 \gamma_E-\log\bigg(\frac{ Q_s^2(x_2)}{4M^2 (1-x_1)} \bigg)+
\psi\left({\frac{7}{2}}\right) \bigg] \;.
\end{equation}
The second part comes again from the finite part of integral in $z$ which is
\be
\int_{x_1}^1 dz \frac{I_{L,\gamma=2}(x_1,z,M^2)-I_{L,\gamma=2}^{(0)}(x_1,z=1,M^2)}{1-z} = \int_{x_1}^1 dz \,\frac{z^3 F_2(\frac{x_1}{z},M^2)-F_2(x_1,M^2)}{1-z} \; .
\ee
We  thus obtain
\begin{equation}
\Delta_{L,4}^{(2)} \; = \; \frac{\alpha_{em}^2\sigma_0}{6\pi^2 M^2}\frac{ 2 }{x_1+x_2} \bigg( \frac{ Q_s^2(x_2)}{4M^2} \bigg)^2 \times \bigg( -\frac{16}{15} \bigg) \int_{x_1}^1 dz \frac{z^3 F_2(\frac{x_1}{z},M^2)-F_2(x_1,M^2)}{1-z}  \;.
\end{equation}
The total contribution to the twist 4 for the longitudinal case is therefore
\begin{equation}
\frac{d^2\sigma^{DY(\tau=4)}_L}{dM^2\,dx_F} \; = \;  \Delta_{L,4}^{(1)}+\Delta_{L,4}^{(2)} \; .
\end{equation}

\bibliographystyle{h-physrev4}
\bibliography{mybib_dy}

\end{document}